# Direct Estimation of Porosity from Seismic Data using Rock and Wave Physics Informed Neural Networks (RW-PINN)

Divakar Vashisth*[1] and Tapan Mukerji[2]

[1]*Department of Energy Science and Engineering, Stanford University, USA,* [2]*Department of Energy Science and Engineering, Stanford University, USA; Department of Geological Sciences, Stanford University, USA; and Department of Geophysics, Stanford University, USA.*

email: [1]*divakar.vashisth98@gmail.com,* [2]*mukerji@stanford.edu*

## Abstract

Petrophysical inversion is an important aspect of reservoir modeling. However due to the lack of a unique and straightforward relationship between seismic traces and rock properties, predicting petrophysical properties directly from seismic data is a complex task. Many studies have attempted to identify the direct end-to-end link using supervised machine learning techniques, but face different challenges such as a lack of large petrophysical training dataset or estimates that may not conform with physics or depositional history of the rocks. We present a rock and wave physics informed neural network (RW-PINN) model that can estimate porosity directly from seismic image traces with no or limited number of wells, with predictions that are consistent with rock physics and geologic knowledge of deposition. As an example, we use the uncemented sand rock physics model and normal-incidence wave physics to guide the learning of RW-PINN to eventually get good estimates of porosities from normal-incidence seismic traces and limited well data. Training RW-PINN with few wells (weakly supervised) helps in tackling the problem of non-uniqueness as different porosity logs can give similar seismic traces. We use weighted normalized root mean square error loss function to train the weakly supervised network and demonstrate the impact of different weights on porosity predictions. The RW-PINN estimated porosities and seismic traces are compared to predictions from a completely supervised model, which gives slightly better porosity estimates but poorly matches the seismic traces, in addition to requiring a large amount of labeled training data. In this paper, we demonstrate the complete workflow for executing petrophysical inversion of seismic data using self-supervised or weakly supervised rock physics informed neural networks.

# Introduction

Seismic data is widely used in exploration and exploitation industry to infer subsurface geophysical and geological properties to eventually delineate an underlying reservoir. The conventional approach for reservoir characterization involves geophysical inversion of seismic data to obtain elastic properties like seismic velocities and densities (Tarantola, 2005; Sen, 2006; Sen and Stoffa, 2013; Biswas *et al.*, 2019; Das *et al.*, 2019) followed by petrophysical inversion with the help of rock physics models obtained from core and well log data to retrieve petrophysical properties like porosities and water saturation (Avseth *et al.*, 2005; Bachrach, 2006; Mavko *et al.*, 2009, Aleardi, 2018; Grana *et al.*, 2021). Several researchers have attempted petrophysical inversion of seismic data using deterministic methods (Angeleri and Carpi, 1982; Dolberg *et al.*, 2000; Bosch, 2004), stochastic methods (Ma, 2002; Eidsvik *et al.*, 2004; Saltzer *et al.*, 2005; Bachrach, 2006) and geostatistical methods (Gonzalez *et al.*, 2008; Bosch *et al.*, 2009a, 2009b; Azevedo and Soares, 2017; Grana *et al.*, 2021). Avseth *et al.* (2005) and Grana *et al.* (2021) have provided a detailed review on different methods and workflows to perform seismic inversion for reservoir characterization.

There is no direct relationship between seismic data and petrophysical properties hence many studies have made use of machine learning methods to approximate the complex nonlinear functional mapping between input (seismic trace) and output (petrophysical properties) without specifying any explicit physical relationships. Some examples include the use of support vector regressor and artificial neural networks (Wong *et al.*, 2002; Gholami and Ansari, 2017; Chaki *et al.*, 2018; Singh *et al.*, 2021). However, these studies are based on supervised learning that requires a large labeled training dataset that is geologically consistent with the target reservoir, which can hinder the use of supervised methods. Das and Mukerji (2020) used convolutional neural networks (CNN) to estimate petrophysical properties directly from seismic data and used very few wells to augment the training data. Jo *et al.* (2021) used ResUNet++ to estimate porosity from spectral decomposed seismic data.

Another shortcoming of solving the inverse problem using purely supervised learning techniques is that the estimates are not guaranteed to be physically or geologically consistent, honoring the different velocity-porosity relations in different depositional environments. To overcome these limitations, we propose a rock and wave physics informed machine learning model inspired by the autoencoder architecture (Goodfellow *et al.*, 2016). This machine learning model learns on its own by maximizing the match between the input and the output. Calderon-Macias *et al.* (1998) used wave physics informed fully connected neural network for automatic normal moveout correction and velocity estimation. Biswas *et al.* (2019) used wave physics guided CNN for seismic impedance inversion while Dhara and Sen (2022) used physics guided autoencoder to perform full waveform inversion. Feng *et al.* (2020) used unsupervised CNN to estimate porosity from poststack seismic data but required a low-frequency prior porosity along with the source wavelet to correctly predict the absolute value of porosity.

In this paper, we estimate porosity directly from normal-incidence seismic image trace using rock and wave physics informed neural network (RW-PINN). We begin with a brief description of our RW-PINN architecture for both self-supervised and weakly supervised cases. Then, in order to demonstrate the efficacy of our proposed architecture, we provide insights into the synthetic data we generated using uncemented sand rock physics model and normal-incidence wave physics model. This is followed by a comparison between results from completely self-supervised and weakly supervised RW-PINN model trained using a few wells with a discussion on the impact of number of wells and weights on normalized root mean square error loss function. Finally, we compare the RW-PINN results with a completely supervised model and discuss the pros and cons of the different approaches.

## Rock and Wave Physics Informed Neural Network (RW-PINN)

The RW-PINN architecture can be sub-divided into two parts: an encoder and a decoder (Figure 1). The encoder is a deep CNN that takes the seismic trace as input and gives a latent vector as output. Then, this output vector acts as input for the decoder where it passes through the rock physics block encapsulating the porosity to velocity relationship, followed by the wave physics block to give normal-incidence reflectivities that are convolved with the source wavelet to output a predicted seismic trace from the decoder. The network learns on its own by minimizing the misfit between the input seismic trace and predicted seismic trace. Once the training is complete, the encoder output vector gets a meaning in the form of petrophysical porosity and the encoder is separated and used to predict porosity directly from the seismic trace. For the completely self-supervised case, root mean square (RMS) misfit is minimized between the input seismic trace and the predicted seismic trace from the decoder.

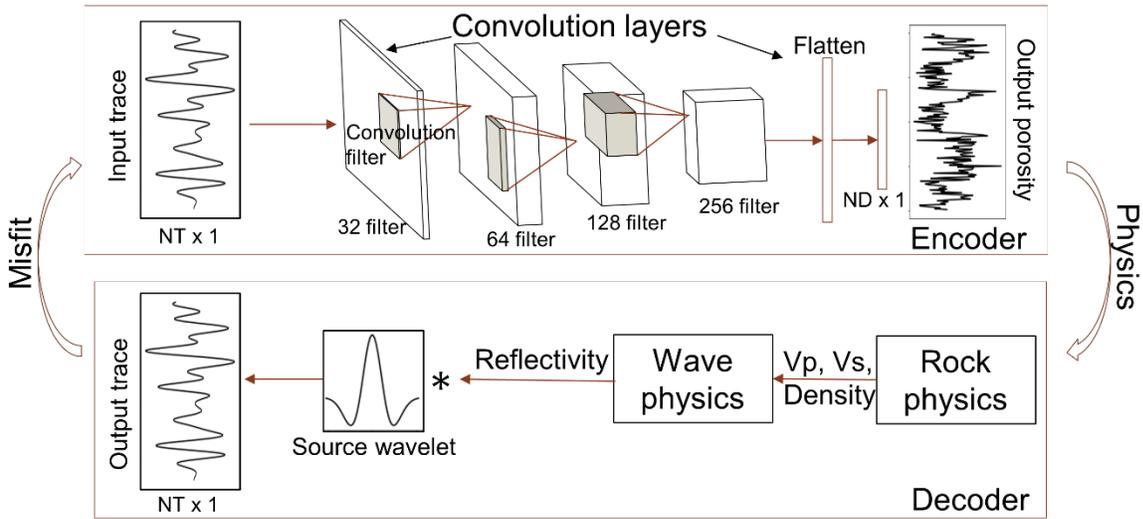

*Figure 1.* Self-supervised RW-PINN architecture, where NT is number of time samples and ND is number of depth samples. Encoder is a deep CNN that is trained to output porosity from a given input seismic trace while decoder has all the physics required to guide the learning of the encoder.

We use He initialization method (He *et al.*, 2015) for weights while bias terms of all the convolution layers are initialized to 0 in the encoder. We use dropout regularization (randomly sets 30% of the input units of a layer to zero at each epoch) only in the last convolution layer to avoid overfitting during training (Srivastava *et al.*, 2014). Rectified linear units (ReLU) activation function (Nair and Hinton, 2010) is used to introduce non-linearity in the CNN network while hyperbolic tangent (tanh) activation function is applied to the CNN output vector to scale the output between -1 to 1 before passing it as input to the decoder, where it is scaled between 0 to 1 to represent porosity. The network learns on its own by backpropagating the misfit between the input seismic and the output seismic trace from the decoder. Tensorflow's (Abadi *et al.*, 2015) automatic differentiation is used to backpropagate the loss through the decoder (i.e. wave and rock physics models) into the encoder. We use a learning rate of 0.0001 to update weights using Adam optimizer (Kingma and Ba, 2014). The RW-PINN model is trained for 200 epochs with a batch size of 128. It should be noted that a number of experiments were performed before finalizing these hyperparameters. Moreover, CNN is used in the encoder as it gave better results in comparison to fully connected neural networks.

*Table 1.* RMS Error in porosity predictions from trained RW-PINN encoder for different weights.

| Architecture | Weight | Training set error | Test set error |
| --- | --- | --- | --- |
| Self-supervised | 0 | 0.06 | 0.06 |
| Weakly supervised | 0.001 | 0.06 | 0.06 |
| Weakly supervised | 0.01 | 0.056 | 0.057 |
| Weakly supervised | 0.1 | 0.050 | 0.050 |

Completely self-supervised RW-PINN does not require any well data to train and is capable of learning on its own but in some scenarios a few wells (porosity data) might be available. In that case, weakly supervised RW-PINN architecture (Figure 2) can be used where the network learns on its own by minimizing the misfit between both the input seismic and output seismic trace from the decoder and the input porosity samples (from well data) and output porosity from the encoder. Specifically, weighted normalized root mean square (NRMS) error between the input and output seismic trace from decoder and input and output porosity from encoder is minimized to train the network. All the hyperparameters for weakly supervised RW-PINN architecture are same as the self-supervised RW-PINN architecture.

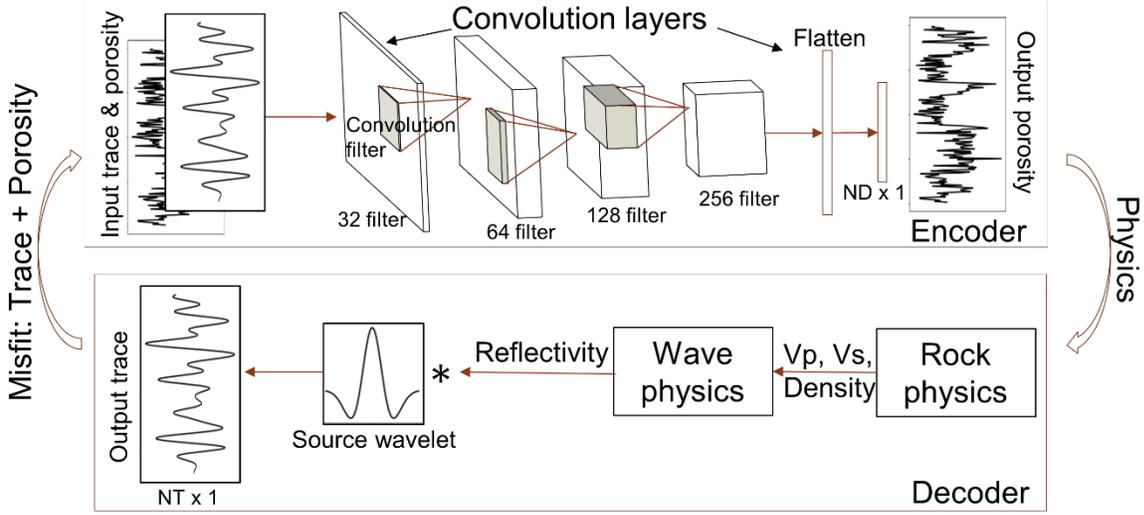

*Figure 2.* Weakly supervised RW-PINN architecture, where NT is number of time samples and ND is number of depth samples. Encoder is a deep CNN that is trained to output porosity from given input seismic trace and porosity logs while decoder has all the physics required to guide the learning of the encoder.

*Table 2.* Comparison of RMS Error in porosity predictions from different architectures.

| Architecture | No. of training wells | Training set error | Test set error |
|---|---|---|---|
| Self-supervised | 0 | 0.06 | 0.06 |
| Weakly supervised | 4 | 0.050 | 0.050 |
| Completely supervised | 2000 | 0.041 | 0.043 |

## Generating the dataset for training and testing RW-PINN

We use the uncemented sand (or soft sand) rock physics model (Mavko *et al.*, 2009) to relate porosities to seismic velocities and densities for water saturated sandstone (80% quartz and 20% feldspar). Following depth to time conversion, normal-incidence reflectivites are computed and convolved with a Ricker wavelet of central frequency 40 Hz to generate seismic traces (Figure 3). Similar steps are followed in the RW-PINN decoder to guide the training of the encoder. The dataset comprises of a total of 2500 porosity vector samples each representing a 200 m thick depth interval sampled at 1 m, and their corresponding seismic traces of which 2000 traces are used for training the RW-PINN (20% of training traces are used for validation) and 500 traces are used for testing. A few of the training and test set traces (input to RW-PINN) are shown in Figure 4a and their corresponding porosity logs (expected output from trained RW-PINN encoder) are shown in Figure 4b.

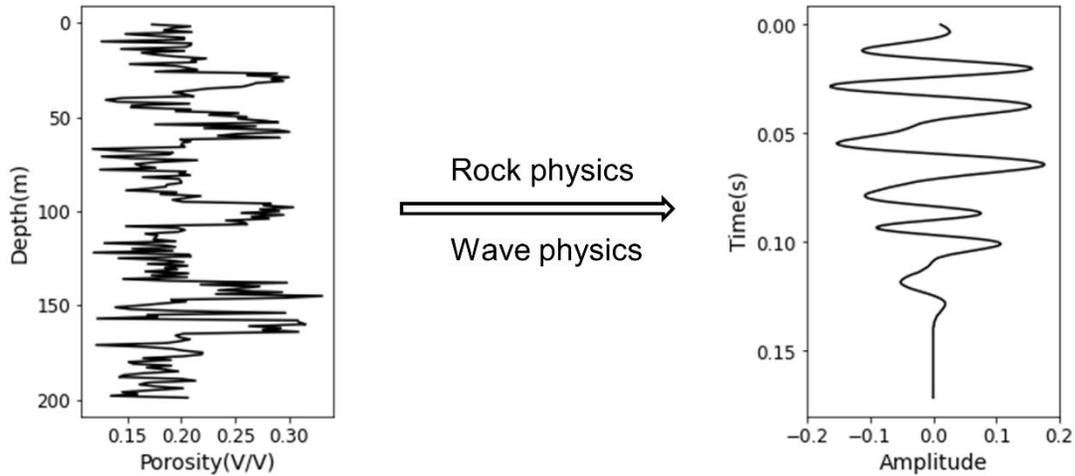

*Figure 3.* Seismic trace (right) generated from porosity log (left) using uncemented sand rock physics model followed by convolving normal-incidence reflectivities with a Ricker wavelet of central frequency 40 Hz.

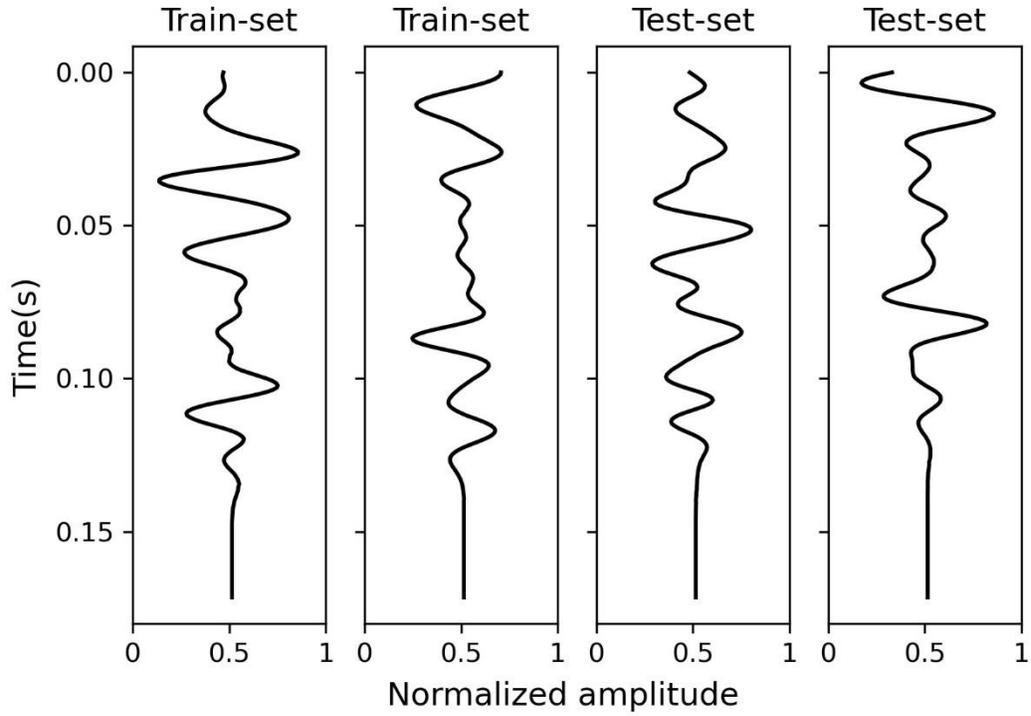

*Figure 4a.* A few training and test set seismic traces (normalized) that are given as input to RW-PINN.

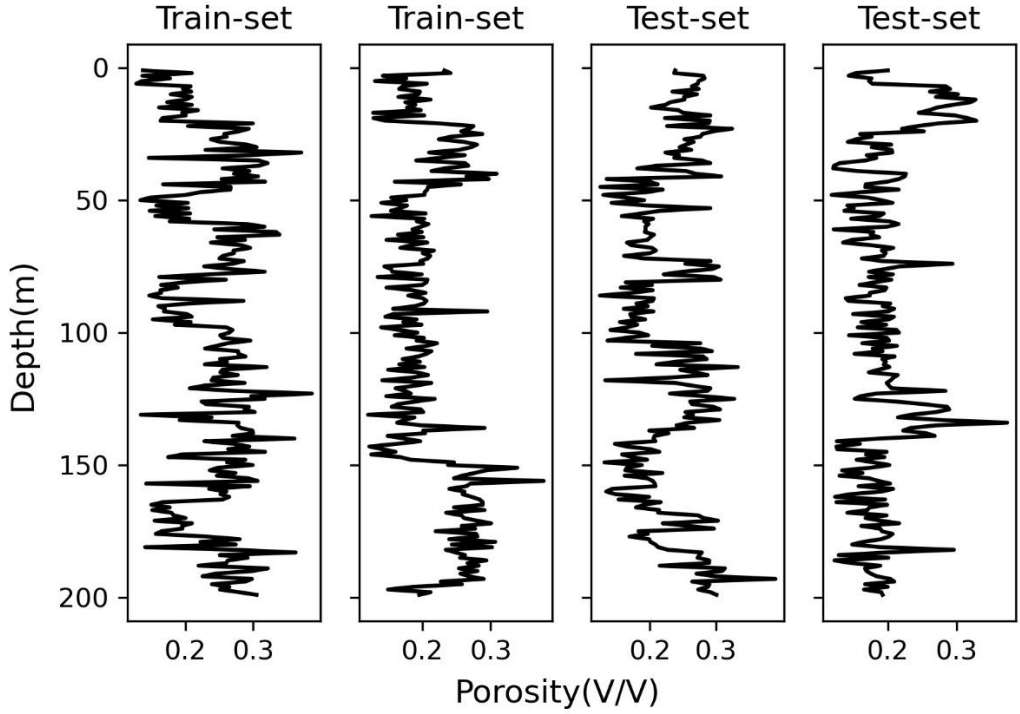

*Figure 4b.* Porosity logs corresponding to seismic traces of Figure 4a that are expected as output from trained RW-PINN encoder.

## Self-supervised, weakly supervised and fully supervised results

In this section we perform tests for three different scenarios: a) fully *self-supervised* with no labeled data, with the network only informed by rock and wave physics; b) *weakly supervised* with a few well porosity logs along with rock and wave physics information; and c) fully *supervised* with a complete labeled training data (seismic trace and porosity log pairs) but no rock physics or wave physics information.

We start with the fully self-supervised case. The input seismic traces and the predicted traces are shown in Figure 5 for the completely self-supervised case. The learning curve is shown in Figure 6. The trained encoder is then separated and used to output porosity samples directly from seismic traces (Figure 7). The average error in porosity prediction is 0.06 for the completely self-supervised RW-PINN. The input and predicted seismic trace (from decoder) show an almost perfect match while some discrepancies are observed in output porosity logs from the trained encoder. This can be attributed to the problem of non-uniqueness where different porosity logs with small scale sub-resolution variations can lead to similar seismic traces for the same rock physics and wave physics models. Some prior information can help to deal with this problem of non-uniqueness. We add prior information in the form of 4 wells to the weakly supervised RW-PINN architecture and allow the network to learn on its own by minimizing the total loss function $E$ shown in equation 1.

$$E = \epsilon_{seismic} + \omega \times \epsilon_{prior\ wells} \qquad (1)$$

where $\epsilon_{seismic}$ corresponds to normalized root mean squared error between the input seismic and output seismic traces from the decoder and $\epsilon_{prior\ wells}$ correspond to normalized root mean squared error between the input prior well porosity logs and output porosity logs (for those same prior wells) from the encoder, and ω is a weighting factor. The NRMS error is the root mean squared error normalized by the mean (expected value) of the training dataset. We experimented with different weights ($\omega$ in equation 1) for the two components of the loss function and got the best porosity predictions (least RMS error) for a weight of 0.1 (Table 1). The output seismic traces from the trained weakly supervised RW-PINN model (with weight of 0.1) for input seismic traces of Figure 4a are shown in Figure 8. The learning curve for the same model is shown in Figure 9. The output porosity logs corresponding to seismic traces of Figure 4a from the trained weakly supervised RW-PINN encoder are shown in Figure 10.

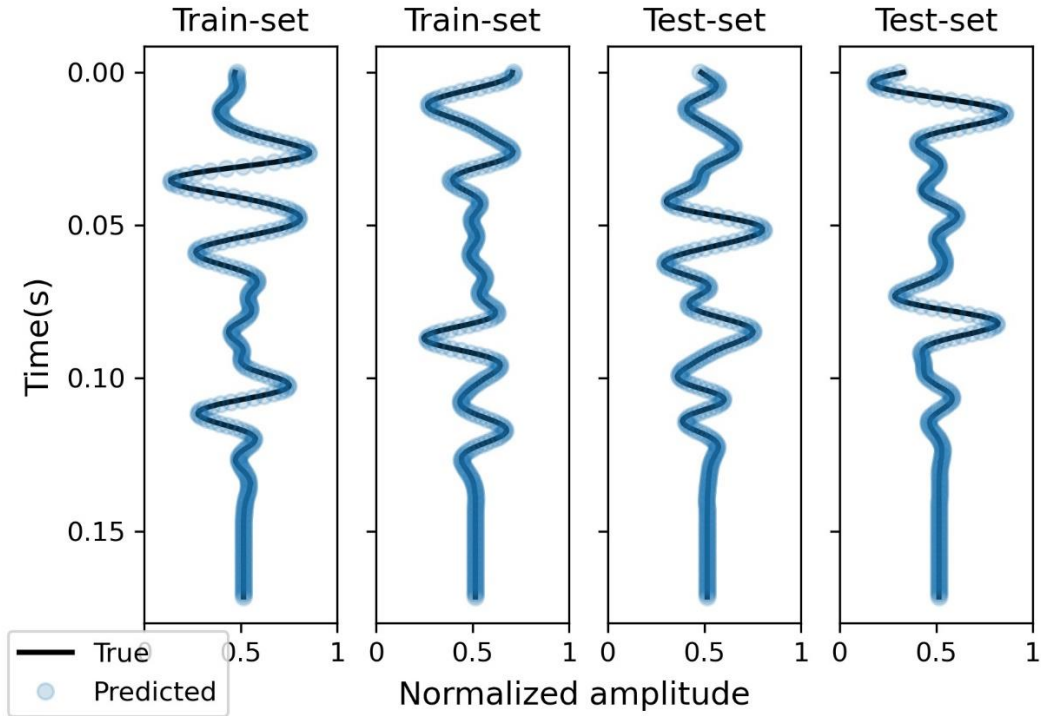

*Figure 5.* Input seismic data (black) to self-supervised RW-PINN and output seismic traces (blue) from trained self-supervised RW-PINN decoder.

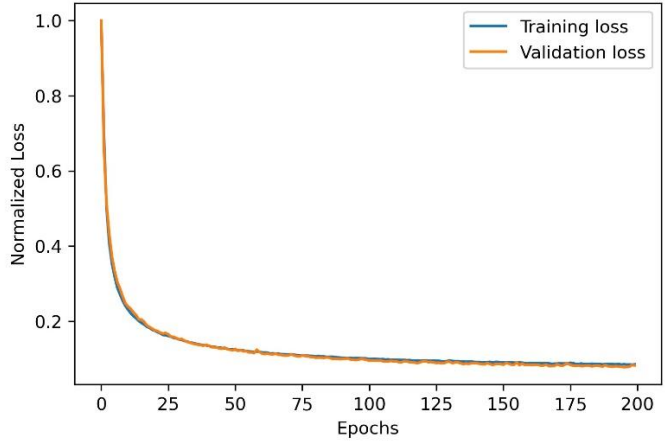

*Figure 6.* Learning curve for self-supervised RW-PINN model.

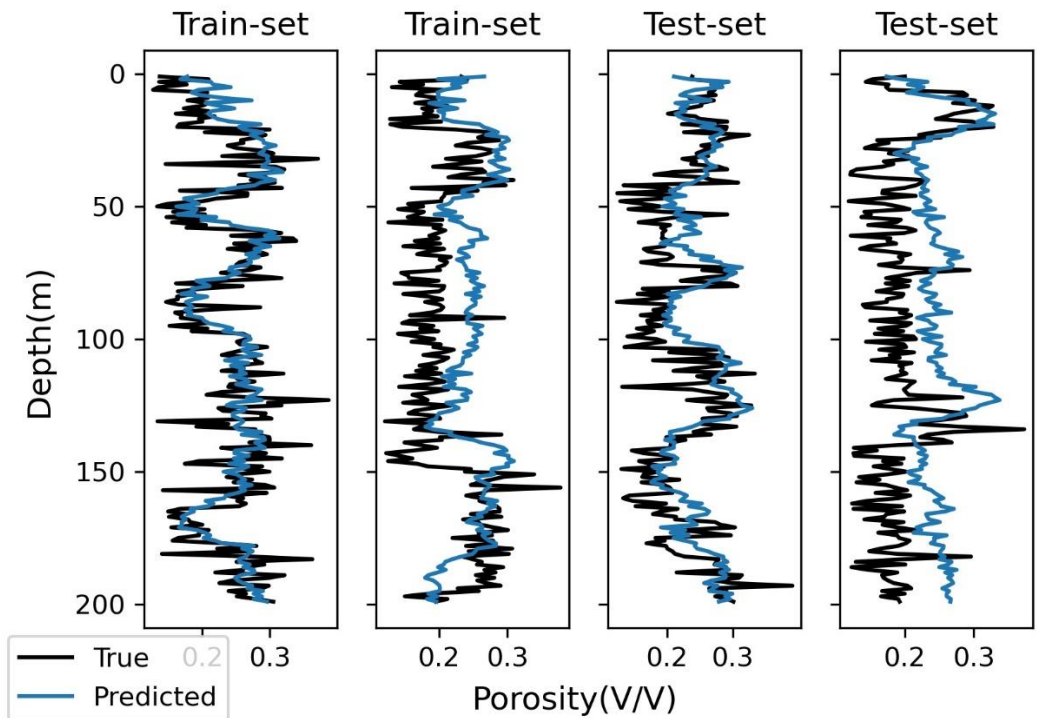

*Figure 7.* True porosity logs (black) corresponding to seismic traces of Figure 4a and predicted porosity (blue) from trained self-supervised RW-PINN encoder.

The output seismic traces (from decoder) match well with the input seismic data (Figure 8), indicating that the physics has been learnt by the weakly supervised RW-PINN model. Moreover, a significant decrease in porosity misfit is observed for the weakly supervised case signifying the impact of adding prior information in the form of 4 wells. It should be noted that the 4 prior wells are different from the 4 porosity logs shown in Figures 4b, 7 and 10.

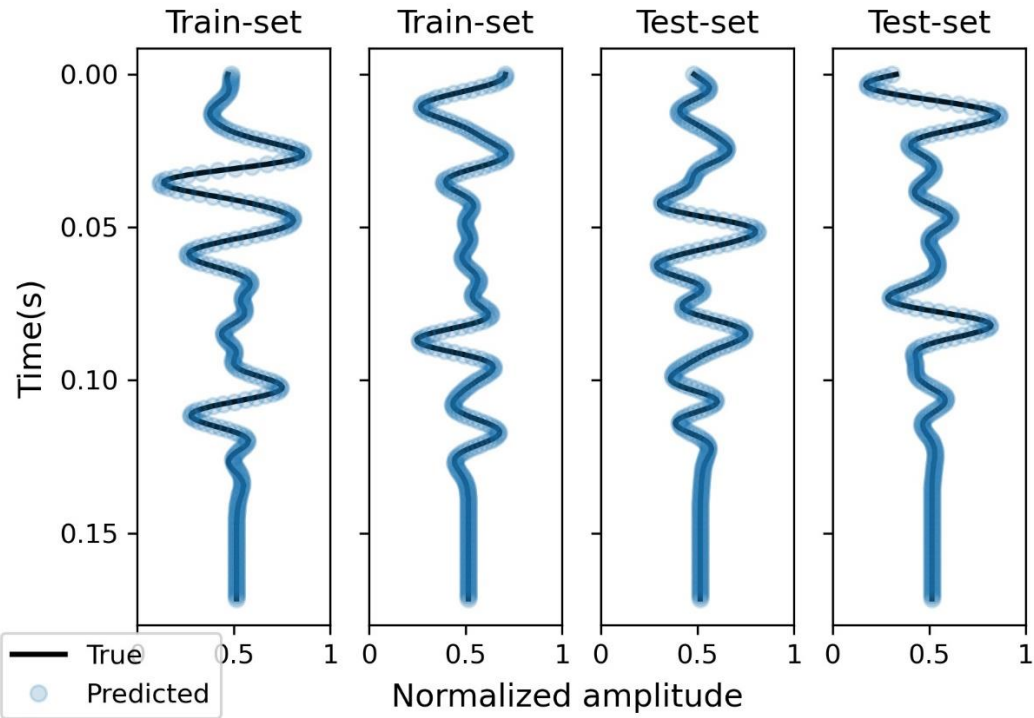

*Figure 8.* Input seismic data (black) to weakly supervised RW-PINN (weight of 0.1) and output seismic traces (blue) from the trained RW-PINN decoder.

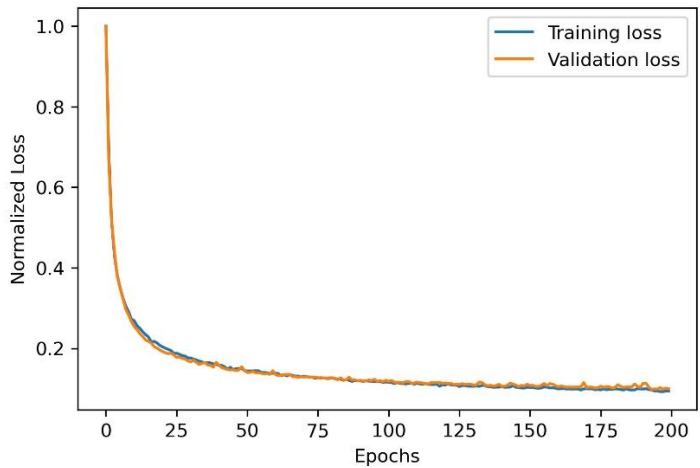

*Figure 9.* Learning curve for weakly supervised RW-PINN model.

*Table 3.* Comparison of RMS Error in seismic predictions from different architectures.

| Architecture | No. of training wells | Training set error | Test set error |
|---|---|---|---|
| Self-supervised | 0 | 0.003 | 0.004 |
| Weakly supervised | 4 | 0.004 | 0.005 |
| Completely supervised | 2000 | 0.032 | 0.033 |

Results are incomplete without the comparison with completely supervised approach that takes seismic data as input and gives porosity logs as output, using a fully labelled training set consisting of pairs of seismic traces and porosity logs. For a fair comparison, we use the same RW-PINN encoder architecture (including the tanh activation function at the end) to perform supervised learning assuming all the porosity logs are available as prior information. Training with 2000 porosity samples leads to overfitting, hence batch normalization (Ioffe and Szegedy, 2015) and dropout regularization is applied to all the convolution layers to prevent overfitting during training. Hyperparameters are kept the same as before in the supervised learning model. The loss function minimized during training is RMS error between the true and estimated porosity from the supervised learning model. The trained supervised model gives porosity logs shown in Figure 11 for input seismic traces of Figure 4a. The predicted porosities match well with the true porosities though the predictions are somewhat smoothed versions of the true values. Overall, porosities estimated from the completely supervised model (trained with 2000 wells) match the true porosities slightly better (RMS error of 0.043) than porosities predicted from weakly supervised RW-PINN encoder (RMS error of 0.05) trained with 4 wells and weight of 0.1 (Table 2). One of the shortcomings of performing petrophysical inversion using supervised learning is that the predicted porosity, when forward modeled to the seismic domain may not give a good match to the seismic trace. To validate this, we generate seismic traces from the estimated porosities given by the supervised model using uncemented sand rock physics and normal-incidence wave physics models (Figure 12) and calculate the mismatch between the true (input) and computed seismic traces from forward modelling. For the sake of comparison, we also calculate the mismatch between input seismic data and output seismic traces from RW-PINN decoder for both self-supervised and weakly supervised cases (Table 3). Clearly, RW-PINN outperforms supervised model (better matching seismic traces) as it is informed by rock physics and wave physics.

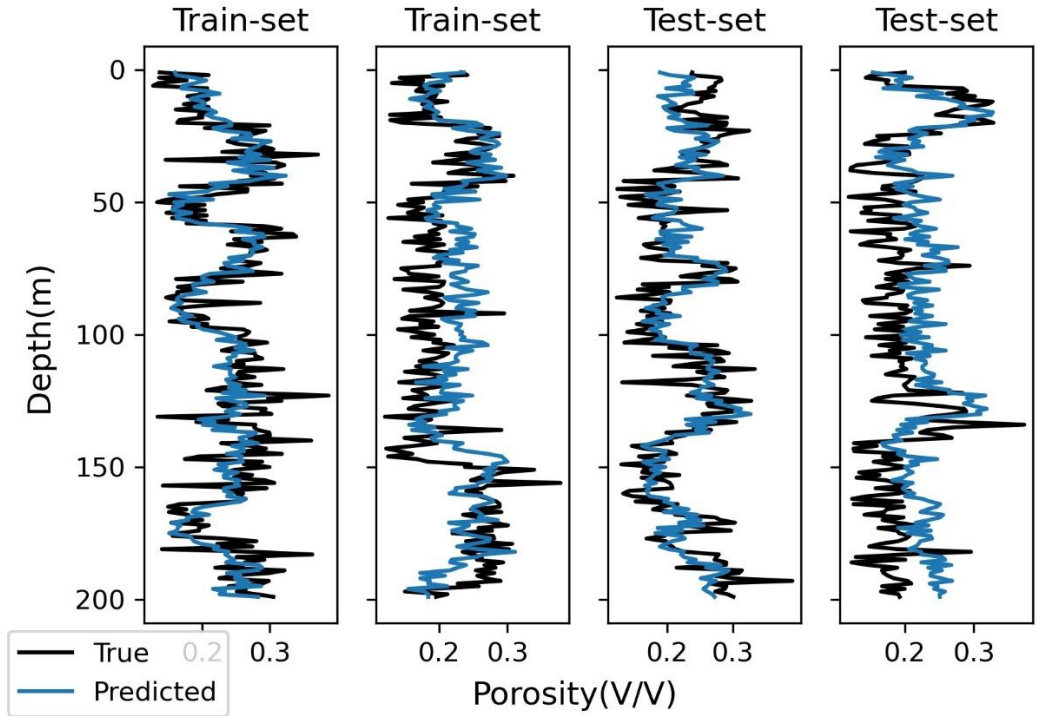

*Figure 10.* True porosity logs (black) corresponding to seismic traces of Figure 4a and predicted porosity (blue) from trained weakly supervised (with weight of 0.1) RW-PINN encoder.

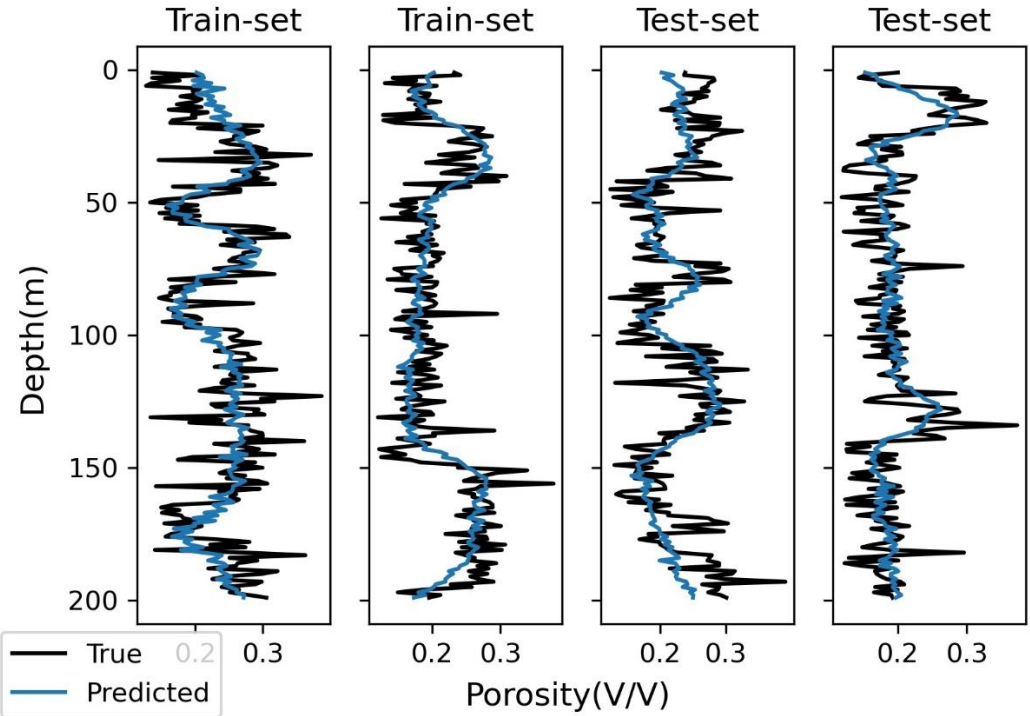

*Figure 11.* True porosity logs (black) corresponding to seismic traces of Figure 4a and predicted porosity (blue) from fully supervised learning model.

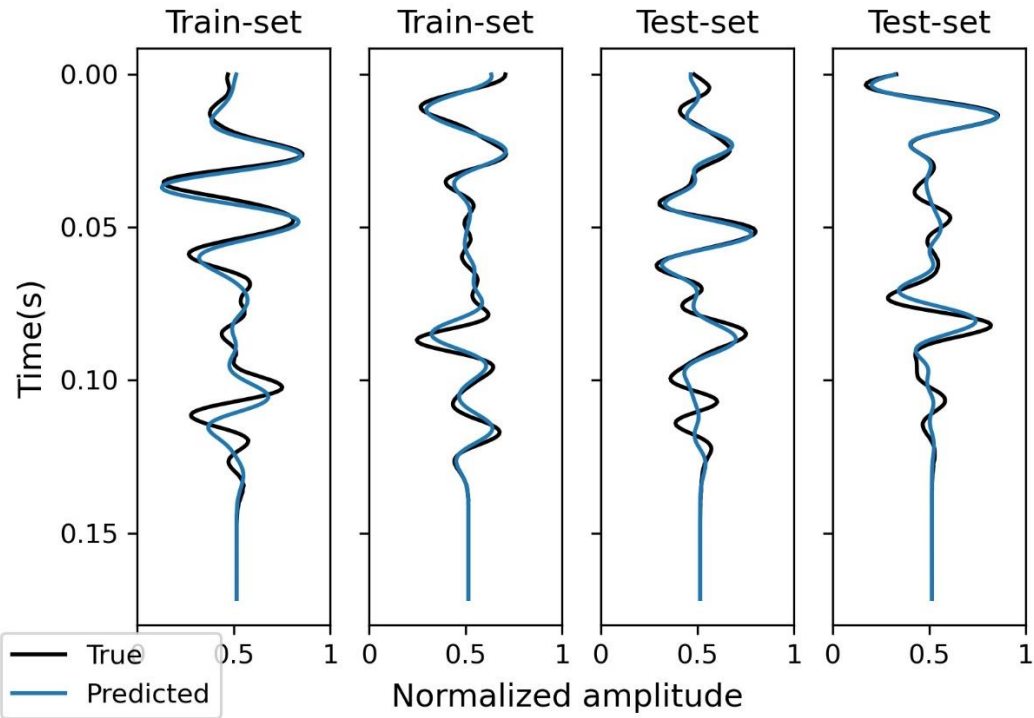

*Figure 12.* Input seismic data (black) and forward modelled seismic traces (blue) using rock and wave physics, from porosities predicted by the supervised model.

## Discussion

In this paper, we estimate porosity directly from seismic traces using rock and wave physics informed neural networks. We have shown the results for a specific velocity-porosity model (the soft sand rock physics model) and normal-incidence wave physics. Other rock physics models such as the stiff-sand model and Berryman's self-consistent model also gave similar results. We also analyzed the effect of including far-angle prestack seismic data in the network but no significant improvements in porosity predictions were observed. We assume only 4 wells are known as prior for the weakly supervised case but adding more wells does not show any major improvement in the porosity predictions. We have not analyzed the impact of partial saturations or the presence of clay on the predictions, which will be part of future work, along with applying the RW-PINN model to field data. The run time for a single epoch was 0.2 seconds for RW-PINN model and 0.1 seconds for the completely supervised model on a single NVIDIA A100 SXM4 GPU. Therefore, the training times for the RW-PINN model and the fully supervised model were 40 and 20 seconds, respectively. For the cases presented in this paper, the run times for RW-PINN and supervised models are almost the same, but when complicated rock physics models are used, the run times for RW-PINN architectures can increase.

## Conclusions

We have introduced the application of rock and wave physics informed neural networks (RW-PINN) to solve petrophysical inverse problems. Specifically, we have used RW-PINN to estimate porosity from seismic traces. For the uncemented sand rock physics model in a completely self-supervised RW-PINN architecture, good porosity predictions (RMS error of 0.06) are observed from the trained encoder but with discrepancies due to the problem of non-uniqueness. We tried to remove these discrepancies by allowing the weakly supervised RW-PINN model to learn with the help of a few prior known porosity logs. Weakly supervised RW-PINN is able to give better porosity predictions with RMS error of 0.05. A completely supervised approach gives better porosity estimates than weakly supervised RW-PINN (in terms of reduced porosity misfit), but the predictions are not consistent with the seismic trace, since the fully supervised model is not informed by the rock and wave physics. Moreover, RW-PINN was almost as fast as the fully supervised model in giving porosity predictions, and has the advantage of not requiring a fully labelled training data. But on the other hand RW-PINN requires coding of the rock physics and wave physics models appropriately to allow for automatic differentiation within TensorFlow. The outputs from the trained RW-PINN encoder (porosities) and decoder (seismic traces) match well with the true data and hence demonstrate the efficacy and effectiveness of the proposed method for solving petrophysical inverse problems for reservoir modelling.

## Acknowledgements

This work is supported by the Stanford Center for Earth Resources Forecasting, and the Stanford Energy Science and Engineering Department. We would like to thank the Dean of the School of Earth, Energy, and Environmental Sciences at Stanford University, S. Graham, for funding. We would also like to thank Mingliang Liu (Stanford University) for helpful discussions.

## Conflicts of Interest

The corresponding author states on behalf of all the authors that there is no conflict of interest.

## Availability of data and materials

Data and materials associated with this research are available and can be obtained by contacting the corresponding author.

Saltzer, R., C. Finn, and O. Burtz, 2005, Predicting VShale and porosity using cascaded seismic and rock physics inversion: The Leading Edge, **24**, 732–736, doi: 10.1190/1.1993269.

Sen, M. K., 2006, Seismic inversion: Society of Petroleum Engineers.

Sen, M.K., and P.L. Stoffa, 2013, Global optimisation methods in geophysical inversion: Cambridge University Press.

Singh, A.P., D. Vashisth, and A. Vishwakarma, 2021, Petrophysical Inversion of seismic dataset using artificial neural networks: Fall Meeting, AGU, Abstracts, S35C-0224.

Srivastava, N., G. E. Hinton, A. Krizhevsky, I. Sutskever, and R. Salakhutdinov, 2014, Dropout: A simple way to prevent neural networks from overfitting: Journal of Machine Learning Research, **15**, 1929–1958.

Tarantola, A., 2005, Inverse problem theory and methods for model parameter estimation: SIAM.

Wong, P. M., A. G. Bruce, and T. D. Gedeon, 2002, Confidence bounds of petrophysical predictions from conventional neural networks: IEEE Transactions on Geoscience and Remote Sensing, **40**, 1440–1444, doi: 10.1109/TGRS.2002.800278.
20